\documentclass[twocolumn,showpacs,amsmath,amssymb,aps,prd,floats]{revtex4-1}
\usepackage{epsfig}
\usepackage{bm}
\usepackage{graphicx}
\usepackage{color}
\usepackage{hyperref}
\usepackage{float}
\usepackage{epstopdf}
\usepackage{wrapfig}

\def\lat{\textit{Fermi}-LAT\/ }

\begin{document}

\title{Neutrino Emission from an Off-Axis Jet Driven by the Tidal Disruption Event AT2019dsg}
\author{Ruo-Yu Liu$^{1,2}$}\email{ryliu@nju.edu.cn}
\author{Shao-Qiang Xi$^{1,2}$}
\author{Xiang-Yu Wang$^{1,2}$}\email{xywang@nju.edu.cn}

\affiliation{$^1$School of Astronomy and Space Science, Nanjing University, Xianlin Road 163, Nanjing 210023, China\\
$^2$Key Laboratory of Modern Astronomy and Astrophysics (Nanjing University), Ministry of Education, Nanjing 210023, China}

\begin{abstract}
Recently, a high-energy muon neutrino event was detected in association with a tidal disruption event (TDE) AT2019dsg at the time about 150\,days after the peak of the optical/UV luminosity. We propose that such a association could be interpreted as arising from hadronic interactions between relativistic protons accelerated in the jet launched from the TDE and the intense radiation field of TDE inside the optical/UV photosphere, if we are observing the jet at a moderate angle (i.e., $\sim 10^\circ - 30^\circ$ ) with respect to the jet axis. Such an off-axis viewing angle leads to a high gas column density  in the line of sight which provides a high opacity for the photoionization and the Bethe-Heitler process, {and allows the existence of an intrinsic long-term X-ray radiation of comparatively high emissivity}. As a result, the cascade emission accompanying the neutrino production, which would otherwise overshoot the flux limits in X-ray and/or GeV band, is significantly obscured or absorbed. Since the jets of TDEs are supposed to be randomly oriented in the sky, the source density rate of TDE with an off-axis jet is significantly higher than that of TDE with an on-axis jet. Therefore, an off-axis jet is naturally expected in a nearby TDE being discovered, supporting the proposed scenario. 
\end{abstract}

\maketitle

\section{Introduction}

A tidal disruption event (TDE) is produced when a star approaches so close to a supermassive black hole (SMBH) that it is torn apart by the tidal force of the SMBH. Part of the stellar debris forms a transient accretion disk around the SMBH, resulting in a luminous panchromatic flare. The majority population of TDEs are thermal TDEs, i.e.,  their X-ray, ultraviolet, and optical radiations have a thermal spectrum. A small fraction of TDEs present nonthermal X-ray emission, implying that powerful relativistic jets are launched in these TDEs \citep{Burrows11, Bloom11, Zauderer11, Cenko12, Brown15}.  It has been suggested that high-energy cosmic rays may be accelerated in the jets of TDEs \citep{Farrar09, Farrar14, ZhangBT17}, and subsequently produce high-energy neutrinos via the $p\gamma$ interactions of the accelerated protons with the intense photon field of the TDE \cite{Wang11_TDE, Wang16, Dai17, Senno17, Lunardini17, Guepin18, Biehl18, Hayasaki19}. Recently, \citet{Stein20} reported, for the first time, the association of a muon neutrino event of energy $E_\nu\gtrsim 200\,$TeV, detected by IceCube on 2019 October 1 (IceCube-191001A), with a radio-emitting TDE (named AT2019dsq) revealed by the Zwicky Transient Facility (ZTF). Radio-emitting TDEs constitute only a small fraction of the bulk population  of TDEs. The chance probability of finding a TDE  as bright as AT2019dsq in the direction of a high-energy neutrino event is  about 0.2\%.

AT2019dsg is located at a redshift of $z=0.051$ or a luminosity distance of $D_L\simeq 230\,$Mpc from Earth \citep{Nicholl19}. Its peak luminosity is in the top 10\% of the 40 known optical TDEs to date\citep{vanVelzen20, Stein20}. By the time of the neutrino detection, which is 150\,days after the luminosity peak in optical/UV (OUV) band, the OUV light curve has reached a plateau, and sustains an OUV luminosity of $L_{\rm OUV}\sim 3\times 10^{43}\,\rm erg~s^{-1}$ with the spectrum being well described by a blackbody photosphere of a near constant temperature of $T_{\rm OUV}\sim 10^{4.59\pm 0.02} {\rm K}$ over time, {implying an OUV photosphere radius of $R_{\rm OUV}=10^{14.1}\,$cm}. The TDE was also bright in the X-ray band with a luminosity of $L_{\rm X}\sim 2.5\times 10^{43}\,\rm erg~s^{-1}$ (in 0.3 -- 10\,keV) discovered at the time of 17\,days after the OUV peak. The X-ray spectrum is consistent with thermal spectrum of a blackbody of temperature $T_{\rm X}=10^{5.9}\,$K, emitted supposedly from a hot accretion disk. {Based on the temperature, the inferred bolometric X-ray luminosity at that time is $7.6\times 10^{43}\,\rm erg~s^{-1}$.} However, this X-ray flux faded extremely rapidly, by a factor of least 50 times over a period of 159 days as measured by Swift/XRT and XMM-Newton \citep{Stein20}. The
rapid decrease of the X-ray flux could be caused by cooling of the newly-formed TDE accretion
disk or increasing X-ray obscuration\citep{Stein20}.  The \textit{Fermi} Large Area Telescope (\lat) did not detect significant signal from the TDE, resulting in an upper limit of $10^{-12}-10^{-11}\rm \, erg~cm^{-2}s^{-1}$ for the flux in $0.1-800\,$GeV averaging over 230\,days after the discovery of the TDE (data analysis is detailed in Appendix; see also Ref.\citep{Stein20}).

According to a 3D fully general relativistic radiation magnetohydrodynamics simulation performed in Ref.\citep{Dai18}, the outflow wind has drastically different density and velocity profiles at different inclination angles. At larger
inclination angles (closer to mid-plane), the outflows are
denser and slower, which would lead to a severe absorption of the X-ray/gamma-ray emission. At lower inclination angles, the outflows are much more dilute (but still as high as $10^{11}\,\rm cm^{-3}$), resulting in a high ionization state so the X-ray photons are efficiently scattered by free electrons; only when observers look down the funnel where the gas density could be as low as $< 10^{9}\, \rm cm^{-3}$, the intrinsic  X-ray emission can be seen.
Therefore, if the rapid decline of the X-ray luminosity seen in AT2019dsg is interpreted as being gradually obscured by the expanding dense outflow, the observers must view the TDE at a moderate inclination angle leastwise ($\theta \gtrsim 10^\circ$).

Recently, Ref.\citep{Winter20} suggested that  a relativistic jet is present  in the TDE  AT2019dsg and emits the neutrino and electromagnetic (EM) radiation along the line of sight. They suggested that a fraction of the X-ray photons are back-scattered by the ionized electrons in the surrounding outflow, providing the target photon field for the production of neutrinos via $p\gamma$ interactions. Ref.\citep{Murase20} proposed non-jetted scenarios for neutrino production, where neutrinos are produced in the core region  (e.g. the hot coronae around an accretion disk). 
In this paper, we propose that the neutrino event and EM radiation of TDE AT2019dsg can be explained self-consistently with a simple one-zone model in the framework of an off-axis jet, in which relativistic protons are accelerated and interact with the intense OUV radiation of the TDE. The rest part of the paper is organized as follows: in Section~\ref{sec:interaction}, we study the relevant interaction processes such as neutrino production and photon absorption. In Section~\ref{sec:result} we show the resulting neutrino spectrum and multi-wavelength flux expected in the model. We discuss and summarize the result in Section~\ref{sec:conclusion}.

\section{Neutrino production and related EM cascade emission}\label{sec:interaction}
Following Ref.\citep{Wang16}, we consider that protons are accelerated in certain dissipation processes in a relativistic jet with a bulk Lorentz factor $\Gamma_j \sim 5$ (corresponding to a jet speed of $v_{\rm j}\simeq 0.98c$ where $c$ is the speed of light). We do not specify the acceleration mechanism here, but for any Fermi-type acceleration mechanism which is common in astrophysical processes, the acceleration timescale can be expressed by a general formula \cite{Aharonian02, Rieger07, Liu17, Lemoine19} $t_{\rm acc}=\eta^{-1} (E_p/\Gamma_jeBc)(c/v_s)^2$ where $\eta(\leq 1)$ is the acceleration efficiency related to the mean free path and the geometry of the acceleration region, $e$ is the electric charge, $B$ is the magnetic field in the acceleration region, and $v_s$ is the speed of the scattering center.
The true energy of the neutrino event could be as high as 1\,PeV \citep{Stein20}, so it probably requires proton to be accelerated at least up to 20\,PeV in the SMBH rest frame (or 4\,PeV in the jet's comoving frame) as the generated neutrino in the $p\gamma$ interaction carries 5\% of parent proton's energy in general. The acceleration should be completed before protons are transported beyond the UV photosphere with the jet because otherwise the collisions between protons and photons will be tail-on dominated, leading to a significant suppression on the interaction efficiency. This requires the acceleration timescale $t_{\rm acc}$ to be smaller than the jet's crossing timescale of the photosphere i.e., $t_{\rm jc}=R_{\rm OUV}/\Gamma_jv_{\rm j}$. Also, the acceleration of protons should  overcome the energy losses due to the $p\gamma$ interactions and the BH pair production on the radiation of the TDE. The energy losses depend on the photon number density of the TDE's radiation, which is given by
\begin{equation}\label{eq:photon_density}
n=\frac{n_0 \epsilon^2}{\exp(\epsilon/kT_{\rm OUV})-1}
\end{equation}
in the SMBH rest frame, where $n_0$ is found by $\int \epsilon nd\epsilon=L_{\rm OUV}/4\pi R_{\rm OUV}^2c$ for the OUV emission. {We take the values of $L_{\rm OUV}$ and $R_{\rm OUV}$ around the neutrino detection time, i.e., $L_{\rm OUV}=3\times 10^{43}\,$erg/s and $R_{\rm OUV}=10^{14.1}\,$cm.} The X-ray photon field follows the same expression except replacing $T_{\rm OUV}$ by $T_{\rm X}$ and $L_{\rm OUV}$ by $f_XL_{\rm X}$. Here $f_X$ represents the fraction of X-ray photons being scattered or isotropized inside the OUV photosphere and we fix this value at 0.1 in the following calculation. {The rapid decline of the X-ray flux could result from two possibilities as mentioned in the previous section: first, if we observe the TDE through the funnel which is transparent to the X-ray emission, the intrinsic X-ray luminosity must have dropped to a negligible level by the time of the neutrino detection; alternatively, if we observe the TDE at a large inclination angle, the decline of the X-ray luminosity could be ascribed to an increasing obscuration by the outflow as the latter is gradually launched from the disk and blocks the observer's line of sight towards the hot inner accretion disk. The intrinsic X-ray luminosity could keep comparatively high in this scenario and we employ $L_X=7.6\times 10^{42}\,$erg/s around the time of the neutrino detection, where we have presumed that the X-ray light curve follows the same behavior of OUV's light cure, which has entered a plateau by the time of the neutrino detection and declined to a level of $\sim 10$\% of the initial value.}  

\begin{figure}[htbp]
\centering
\includegraphics[width=1\columnwidth]{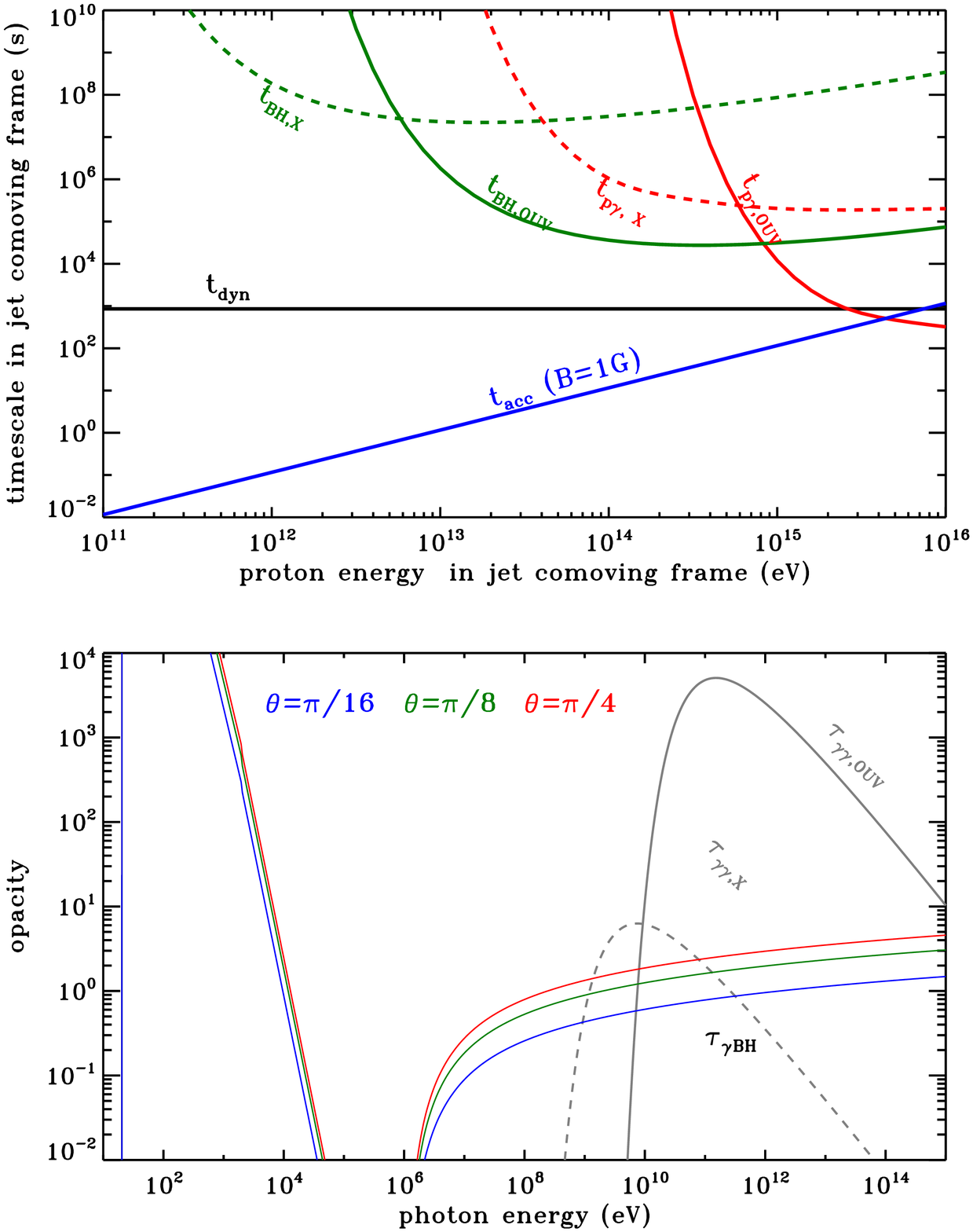}
\caption{{\bf Upper panel:} Relevant timescales for protons. All the timescales are measured in the {jet's comoving frame}. {\bf Lower panel:} Opacities of different processes for photons. {The photon energy is measured in the SMBH rest frame.} The solid/dashed grey curve represents the $\gamma\gamma$ opacity caused by the OUV emission and isotropized X-ray emission. The curves marked with $\tau_{\rm ion}$ and $\tau_{\gamma\rm BH}$ show the opacities of the photoionization and the BH pair production for gamma rays, respectively. The blue, green and red curves represent different viewing angles as labelled. Note that for the on-axis case ($\theta=0$), the gas density is significantly lower than the off-axis cases, so the photoionization opacity and the BH opacity are negligible.}
\label{fig:timescales}
\end{figure}

Note that, in Eq.~(\ref{eq:photon_density}), we assume that the proton acceleration and interaction region is located around the OUV photosphere. As we mentioned above, a larger radius is not favored for the neutrino production. On the contrary, one could consider a smaller radius which would benefit the neutrino production. The energy loss timescale of protons due to the $p\gamma$ interaction ($t_{p\gamma}$) and the BH pair production ($t_{\rm BH}$) are treated following the semi-analytic method developed by \citep{Kelner08}. Relevant timescales of protons as functions of proton energy are shown in the upper panel of Fig.~\ref{fig:timescales}. We consider the most efficient acceleration case (i.e., $\eta\sim 1$ and $v_s\sim c$) and a conservative lower limit for the magnetic field {$B\simeq 1\,$G} can be obtained from the requirement $t_{\rm acc}<(t_{p\gamma}^{-1}+t_{\rm BH}^{-1})^{-1}$ for 20\,PeV proton.

By comparing the jet crossing timescale and the $p\gamma$ interaction timescale, we find that the OUV radiation field leads to an almost full $p\gamma$ interaction efficiency above $\sim 2\,$PeV in the jet's comoving frame or $\sim 10$\,PeV in the SMBH rest frame (corresponding to neutrino energy above $\sim 0.5\,$PeV). High-energy gamma rays, electrons and positrons will be produced associately with neutrinos in $p\gamma$ interactions. High-energy gamma rays will be absorbed by the OUV and/or the isotropized X-ray radiation field of the TDE via the $\gamma\gamma$ annihilation and produce high-energy electron/positron pairs, triggering the EM cascades and depositing their energies into kev -- GeV band via the synchrotron radiation and the IC radiation. One of the generated electron and positron in the annihilation will bring most energy of the high-energy photon when the center-of-momentum energy far exceeds the rest energy of an electron (i.e., $\sqrt{E_\gamma\epsilon} \gg m_ec^2$). The high-energy electron or positron will pass most of its energy to one of the TDE photons via the inverse Compton (IC) scattering in the deep Klein-Nishina regime, and regenerate a high-energy photon. Such a $\gamma-e-\gamma$ cycle, which is also called ``EM cascade'', will proceed several times until the pair production opacity of the new generated photons falls below unity, which occurs around GeV energy if the X-ray radiation presents or around 10\,GeV energy if not (see the lower panel of Fig.~\ref{fig:timescales}). 
Eventually, the high-energy EM particles deposit their energies into keV -- GeV band via the synchrotron radiation and the IC radiation.
Thus the keV flux upper limit measured by XMM-Newton and the GeV flux upper limit measured by \lat could constrain the model, especially the viewing angle. This is because the viewing angle determines the gas column density of the TDE outflow in the line of sight, which is crucial for the absorption processes such as the photoionization and the BH process. {We follow Ref.\citep{Ginzburg64} for the cross section of the former process and Ref.\citep{Chodorowski92} for that of the latter process.} In the lower panel of Fig.~\ref{fig:timescales}, we show the opacities of various attenuation processes for photons with different viewing angles \footnote{We normalize the simulated density profile to the case with $M_{\rm BH}=10^7\,M_\odot$. Since we set the dissipation region around the OUV photosphere, we calculate the  atom column density from $2R_{\rm OUV}$ up to a sufficiently large radius, e.g., 3000$R_{\rm OUV}$, where the simulation reaches. We find that the column density of the outflow are $4.2\times 10^{25}\rm cm^{-2}$, $8.7\times 10^{25}\rm cm^{-2}$ and  $1.2\times 10^{26}\rm cm^{-2}$ for $\theta=\pi/16, \pi/8$, and $\pi/4$ respectively.}, based on the gas density profile simulated by Ref.\citep{Dai18}. {The photoionization opacity is larger than unity at $\lesssim 10\,$keV and quickly drops to zero at $13.6\,$eV due to the threshold of the photoionization of hydrogen atoms. Therefore low energy UV photons can escape from
the system while X-rays are significantly attenuated.} {Here we do not consider the Thomson scattering opacity for two reasons: first, the Thomson scattering opacity depends on the ionization state of the gas since only free electrons provide the opacity, which would then require a sophisticated treatment of the radiative transfer process in the disk and outflows, which is complex and beyond the scope of this work. On the other hand, the scattering process mainly operates on the photons below MeV due to the Klein-Nishina effect, and it won't influence our conclusion whether we take in account this process or not, as will be shown in the next section. In the case that observers look down the funnel region (i.e., $\theta\gtrsim 0$), the gas density in the line of sight is low and we neglect the gas column density for simplicity.} Note that the $\gamma\gamma$ annihilation opacity does not depend on the viewing angle because the target photon field is the isotropic/isotropized OUV and X-ray radiation of the TDE.  
 
\section{Results}\label{sec:result}
We carry out a numerical calculation, as detailed in the Appendix, to obtain the spectrum of muon neutrinos and the accompanying EM cascades expected in our proposed model. We assume that the injection proton spectrum follows the form $dN/dE_p\propto E_p^{-2}\exp\left(-E/E_{p,\rm max}\right)$ for $E_p \geq \Gamma_j \times 1\,$GeV. Since the neutrino flux should not drop significantly by the time of the detection of IceCube-191001A, we consider that the system has reached a quasi-steady state, and take a constant injection luminosity of relativistic protons in the jet, i.e., $L_{p,\rm j}=3\times 10^{44}\rm erg~s^{-1}$ {(measured in the SMBH rest frame)} which is comparable to the peak luminosity of the TDE.
Four different viewing angles, i.e., $\theta=0,\pi/16,\pi/8,\pi/4$ are taken into account in the calculation. Given a typical half opening angle $\theta_j=10^\circ$ for the jet, the observer would see an off-axis jet with the latter three $\theta$. Similar to the assumption used in Fig.~\ref{fig:timescales}, we consider 10\% of the X-ray emission from the accretion disk is isotropized (i.e., $f_{\rm X}=0.1$) providing a target photon field for relevant particle interactions in the cases of $\theta \neq 0$, while we assume the intrinsic X-ray emission has already decayed to a negligible level in the case of $\theta=0$ at a time much earlier than the neutrino detection.

The intrinsic emissivity of neutrino and the EM cascade is more or less the same in the jet's comoving frame regardless of the viewing angle if we ignore the difference caused by the X-ray photon field between the $\theta=0$ case and the $\theta\neq 0$ cases. However, the Doppler effect makes a profound impact on the observed flux. The Doppler factors are $\delta_D=\left[\Gamma_j(1-\beta_j\cos\theta)\right]^{-1}=10, 5.1, 2.1, 0.65$, respectively, for $\theta=0, \pi/16, \pi/8, \pi/4$  with $\Gamma_j=5$. Let us denote the intrinsic differential luminosity of neutrino/EM emission at energy $E'$ in the jet's comoving frame by $L_{\nu/\gamma}'(E')$. Note that, the hadronic emission would fade rapidly once the accelerated protons are transported beyond the photosphere while fresh relativistic protons are continuously injected into the photosphere. Consequently, observers would see that the emission always come from the region within the photosphere as if the emitting region were stationary during the observational period, which is much longer than the  jet's crossing time in the SMBH rest frame $\Gamma_jt_{\rm jc}\sim 4000\,$s. Thus, the flux seen by the observer is (de)boosted by a factor of $\delta_D^3$ for an viewing angle $\theta>\theta_j$ (i.e., off-axis), i.e., 
\begin{equation}\label{eq:nuflux}
F_\nu(\delta_DE')=\frac{\delta_D^3L_\nu'(E')}{4\pi D_L^2}
\end{equation}
for neutrino and
\begin{equation}\label{eq:gammaflux}
F_\gamma=\frac{\delta_D^3L_\gamma'(E')}{4\pi D_L^2}f_{\rm \gamma\gamma}\exp (-\tau_{\rm ion})\exp (-\tau_{\gamma,\rm BH}).
\end{equation}
for photon. The factors $f_{\gamma\gamma}=[1-\exp (-\tau_{\gamma\gamma})]/\tau_{\gamma\gamma}$ in Eq.~(\ref{eq:gammaflux}) is the fraction of the emission that can escape the $\gamma\gamma$ absorption. $\exp (-\tau_{\rm ion})$ and $\exp (-\tau_{\gamma, \rm BH})$ account for the absorption of photons via the photonionization and via BH process in the dense outflow. For the fluxes in the case of $\theta=0$ (i.e., when viewing the jet on-axis), we need to replace $\delta_D^3$ in Eq.~(\ref{eq:nuflux}) and Eq.~(\ref{eq:gammaflux}) with $\delta_D (\theta_j^2/4)^{-1}$. Finally, the gamma-ray absorption during the propagation in the intergalactic space is considered following the model of the extragalactic background light given by \citep{Finke10}.

\begin{figure}[htbp]
\centering
\includegraphics[width=1\columnwidth]{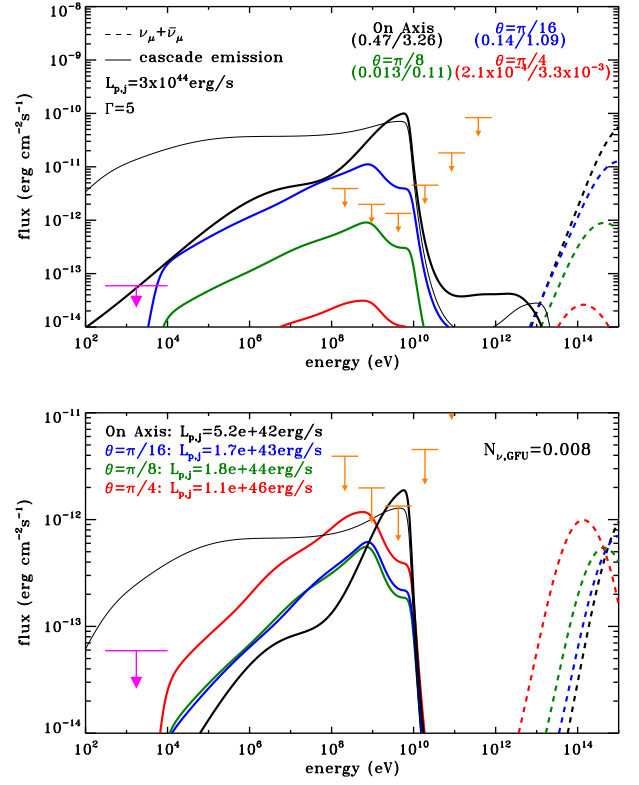}
\caption{{\bf Upper panel:} Expected spectrum of the neutrino emission and EM cascade with different viewing angles (black curves for on-axis case or $\theta=0$; blue, green and red curves for off-axis case with $\theta=\pi/16,\pi/8,\pi/4$ respectively), for $B=1\,$G. The numbers marked under the values of $\theta$ represent the expected muon (including anti-muon) neutrino event number $N_{\nu_{\mu}}$ detected in $(0.2-1)$\,PeV in 150\,days with the GFU/PS effect area of IceCube. The $3\sigma$ upper limits of the TDE's GeV flux averaging over 230\,days after the TDE are shown in orange bars with arrows. The $3\sigma$ upper limit in X-ray band ($0.3-10$\,keV) measured by XMM-Newton is shown as magenta bars and arrows. The thin solid black curves show the expected flux in the on-axis case for $B=500\,$G. {\bf Lower panel:} Same with the upper panel but normalizing $N_{\nu_{\mu}}$ to 0.008 which is the smallest reasonable event number for the detection of a single muon neutrino event in association with AT2019dsg among all seventeen TDEs detected by ZTF.}
\label{fig:spec}
\end{figure}

The result is shown in Fig.~\ref{fig:spec}. In the upper panel, we can see significant differences among four viewing angles considered here. The differences in peak energies and peak fluxes in the neutrino spectra are caused by different Doppler factors due to different viewing angle. Since the X-ray emission of the TDE has to be very weak in the on-axis case, the neutrinos are produced only on the OUV radiation and hence the spectral shape is narrower and GeV gamma rays are little absorbed (see also the lower panel of Fig.~\ref{fig:timescales}). The X-ray and GeV flux upper limits measured by, respectively, XMM-Newton and \lat pose strong constraints on the viewing angle, {and are inconsistent with the $\theta=0$ case and $\theta=\pi/16$ case. Note that the resulting X-ray fluxes in three $\theta \neq 0$ cases do not violate the X-ray upper limits, validating the simplification of neglecting the Thomson scattering process which would otherwise only further reduce the X-ray flux.} The expected $\nu_\mu+\bar{\nu}_\mu$ event number in $(0.2-1)\,$PeV, assuming the emission lasts for 150 days, is obtained by convolving the neutrino flux with the gamma-ray follow-up (GFU) and point-source (PS) effective area of IceCube \citep{Blaufuss19} respectively, as labelled in the upper panel of Fig.~\ref{fig:spec}. The neutrino event number expected with the GFU selection is $0.008\lesssim N_{\nu_\mu} \lesssim 0.76$ for AT2019dsg considering that IceCube-191001A is the only event triggered the IceCube alert among all seventeen TDEs detected by ZTF \citep{Stein20}. On the basis of this analysis, the case of $\theta=\pi/8$ can successfully explain the neutrino--TDE  association and in the meantime conform with the X-ray and GeV upper limits, under the employed parameters. {The employed magnetic field (i.e., $B=1\,$G) corresponds to a magnetic luminosity of $L_B \simeq \pi \theta^2R_{\rm ouv}^2 \Gamma_j^2c (B^2/8\pi) \simeq 5\times 10^{37}(B/1\rm G)^2(\theta/10^\circ)^2\rm erg~s^{-1}$ which is energetically reasonable since it is much smaller than the jet's kinetic luminosity. It allows us to employ a larger magnetic field and hence a less extreme condition of particle acceleration with $\eta \ll 1$.}

It should be noted that the jet's relativistic proton luminosity is not necessarily comparable to the TDE's peak luminosity. In fact, it is rather a free parameter in the model, and the resulting neutrino flux and EM emission flux are linearly proportional to it. We then normalize the theoretical neutrino event number expected in the GFU effective area to 0.008 by rescaling the proton luminosity. As we can see in the lower panel of Fig.~\ref{fig:spec}, the expected flux in the on-axis case still overshoots the \lat $3\sigma$ upper limits. The GeV emission is radiated through the IC process of the cascaded electrons, so it could be suppressed if a large magnetic field is employed. To achieve this, the magnetic field energy density in the comoving frame of the jet should be higher than that of the radiation field, leading to a requirement of $B \gtrsim 1000\,$G. {Indeed, we find that if we use $B=500\,$G, the GeV gamma-ray flux can be reduced to a level marginally consistent with the $3\sigma$ upper limits of \lat.} However, the synchrotron radiation in this case is significantly enhanced and overshoots the X-ray upper limit in the on-axis case, as shown in the thin solid black curves in Fig.~\ref{fig:spec}. Moreover, primary electrons are likely accelerated with protons as well. The accelerated electrons also generate multi-wavelength EM emission via the synchrotron radiation and the IC radiation, consequently aggravating the inconsistency with these upper limits. {Also, given the magnetic luminosity being $L_B \simeq 5\times 10^{43}(B/1000{\rm G})^2\rm erg~s^{-1}$,  the magnetic luminosity is much higher than the proton luminosity for $B\gtrsim 1000$G, which may be at odds with the blazar modelings and theoretical expectation.} Therefore the on-axis jet scenario is less favored.
In the case of $\theta=\pi/4$, the neutrino flux is severely Doppler de-boosted and an unrealistically large proton luminosity of the jet is needed to explain the detection. To conclude, the viewing angle should be sufficiently large ($\gtrsim 10^\circ$) to make the EM emission being obscured, while on the other hand it should remain moderate to avoid neutrino flux being de-boosted (i.e., $\delta_D< 1$). We then expect $10^\circ \lesssim \theta \lesssim 30^\circ$ to be a favorable range of the viewing angle, for a typical Lorentz factor of $5-10$ for the TDE jet.

\section{Discussion and Conclusion}\label{sec:conclusion}
It is speculated that jets in TDEs, if not powerful enough to break out of the dense envelope formed by unbound stellar material, could be choked inside the envelop, finally dissipating all the jet's energy to the cocoon\citep{Wang16, Senno17}. The proposed model in this paper for the neutrino event does not depend on whether the jet can break out the envelop or not, since neutrino emission can be observed in both cases. If the jet is choked, however, the dense envelop could also hide the EM emission even if we observe the jet on-axis \citep{Wang16}. Therefore, the on-axis, choked jet case is probably not constrained by the X-ray and GeV upper limits and remains a possible solution. Nevertheless, since the jets of TDEs are supposed to be randomly oriented in the sky, we would naturally expect the presence of an off-axis jet (no matter choked or not) rather than an on-axis jet in a nearby TDE being discovered. 

On the other hand, the multi-wavelength observation immediately after the jet-breakout may be useful to diagnose whether the jet is successfully launched on-axis. If the jet breaks out the envelope, the jet will propagate in the interstellar medium and produce an external shock. Electrons will be accelerated in the external shock and give rise to non-thermal afterglow emission. For an off-axis observer, as the jet is decelerating,  beaming angle is widening, so the observer would be able to see  a rising non-thermal afterglow emission. However, this afterglow emission is comparatively weak due to being less Doppler boosted or even deboosted, so it may be hard to distinguish it from the emission produced by other TDE components. On the other hand, if we observe a successful jet on-axis, the early afterglow emission would be very bright due to the relativistic boosting. Thus, the on-axis, successful jet scenario may be also disfavored by nondetection of multi-wavelength afterglow of AT2019dsg (see also Ref.\cite{Murase20}).


To conclude, we showed that the association between IceCube-191001A and TDE~2019dsg may be interpreted by an off-axis jet model. The favored viewing angle with respect to the jet axis is $10^\circ-30^\circ$ with which the neutrino flux would not be Doppler de-boosted and in the mean time the accompanying X-ray and GeV gamma-ray emission can be absorbed by the slow, dense outflow in the line of sight. TDEs of off-axis jets may potentially make a considerable contribution to the diffuse high-energy neutrino background and this is to be studied further in the future.

\section*{Acknowledgements}
We would like to thank the anonymous referee for the constructive comments. This work is supported by NSFC grants 11625312 and
11851304, and the National Key R \& D program of China under the grant 2018YFA0404203.

\appendix
\section*{Calculation of the Neutrino and the EM cascade Emission}
The quasi-steady state spectrum of relativistic protons can be given by
\begin{equation}
\frac{dN_p}{dE_p}=\frac{d\dot{N}_p}{dE_p}\left(t_{\rm jc}^{-1}+t_{p\gamma}^{-1}+t_{p,\rm BH}^{-1}\right)^{-1}
\end{equation}
We then deal with the calculation in the jet's comoving frame and covert the relevant quantities to the comoving frame, i.e., the proton energy $E_p'=E_p/\Gamma_j$, the proton spectrum $E_p'dN_p'/{dE_p'}=\Gamma_j E_pdN_p/dE_p$, the differential density of the target photon field $n'(\epsilon')=(n_0/\Gamma_j^2)\epsilon'^2/(\exp(\epsilon'/\Gamma_j kT)-1)$, and the timescale of the $i$ process $t'_i=t_i/\Gamma_j$,  while the opacities are Lorentz invariants.

The generated spectrum of gamma rays, electrons/positrons (hereafter, for simplicity we do not distinguish positrons from electrons as the difference between these two particles are not relevant for this study), and neutrinos from the $p\gamma$ process and the BH process are calculated following the semianalytical method developed by Ref.\cite{Kelner08}, which is denoted by
\begin{eqnarray}
\frac{d\dot{N}_\gamma'}{dE_\gamma'}\Big \vert_{p\gamma}=\mathcal{F}_{p\gamma}^{\gamma}\left\{\frac{dN_p'}{dE_p'}, n'(\epsilon ') \right\}\\
\frac{d\dot{N}_e'}{dE_e'}\Big \vert_{p\gamma}=\mathcal{F}_{p\gamma}^e\left\{\frac{dN_p'}{dE_p'}, n'(\epsilon ') \right\}\\
\frac{d\dot{N}_\nu'}{dE_\nu'}\Big \vert_{p\gamma}=\mathcal{F}_{p\gamma}^{\nu}\left\{\frac{dN_p'}{dE_p'}, n'(\epsilon ') \right\}\\
\frac{d\dot{N}_e'}{dE_e'}\Big \vert_{\rm BH}=\mathcal{F}_{{\rm BH}}^e\left\{\frac{dN_p'}{dE_p'}, n'(\epsilon ') \right\}
\end{eqnarray}
where $\mathcal{F}_{p\gamma}$ and $\mathcal{F}_{\rm BH}$ are the operators to derive the spectra of secondary particles generated in the $p\gamma$ interactions and the BH processes. The generated neutrinos will escape the source directly while gamma rays and electrons are subject to further interactions inside the source.

The intense radiation field of the TDE will absorb high energy gamma rays and produce an electron-positron pair. The fraction of high-energy gamma rays that can escape the source is $f_{\rm esc}=[1-\exp (-\tau_{\gamma\gamma})]/\tau_{\gamma\gamma}$. The total spectrum of gamma-ray photons containing in the source can then be given by
\begin{equation}\label{eq:Ngamma}
\frac{dN_\gamma}{dE_\gamma}=\frac{d\dot{N}_{\gamma'}}{dE_\gamma'}\left(t_{\rm lc}^{'-1}+t_{\gamma\gamma}^{'-1}\right)^{-1}
\end{equation}
where $t'_{\rm lc}=R_{\rm OUV}/\Gamma_jc$ is the light crossing time of the OUV photosphere. Similarly, the electron spectrum generated by the $\gamma\gamma$ annihilation can be given by
\begin{equation}
\frac{d\dot{N}_e'}{dE_e}\Big \vert_{\gamma\gamma}=\mathcal{F}_{\gamma\gamma}^e\left\{\frac{dN_\gamma'}{dE_\gamma'}, n'(\epsilon ') \right\}
\end{equation}
where $\mathcal{F}_{\gamma\gamma}$ follows the expressions shown in Ref.\cite{Aharonian83_epp}.

\begin{figure}[htbp]
\centering
\includegraphics[width=0.9\columnwidth]{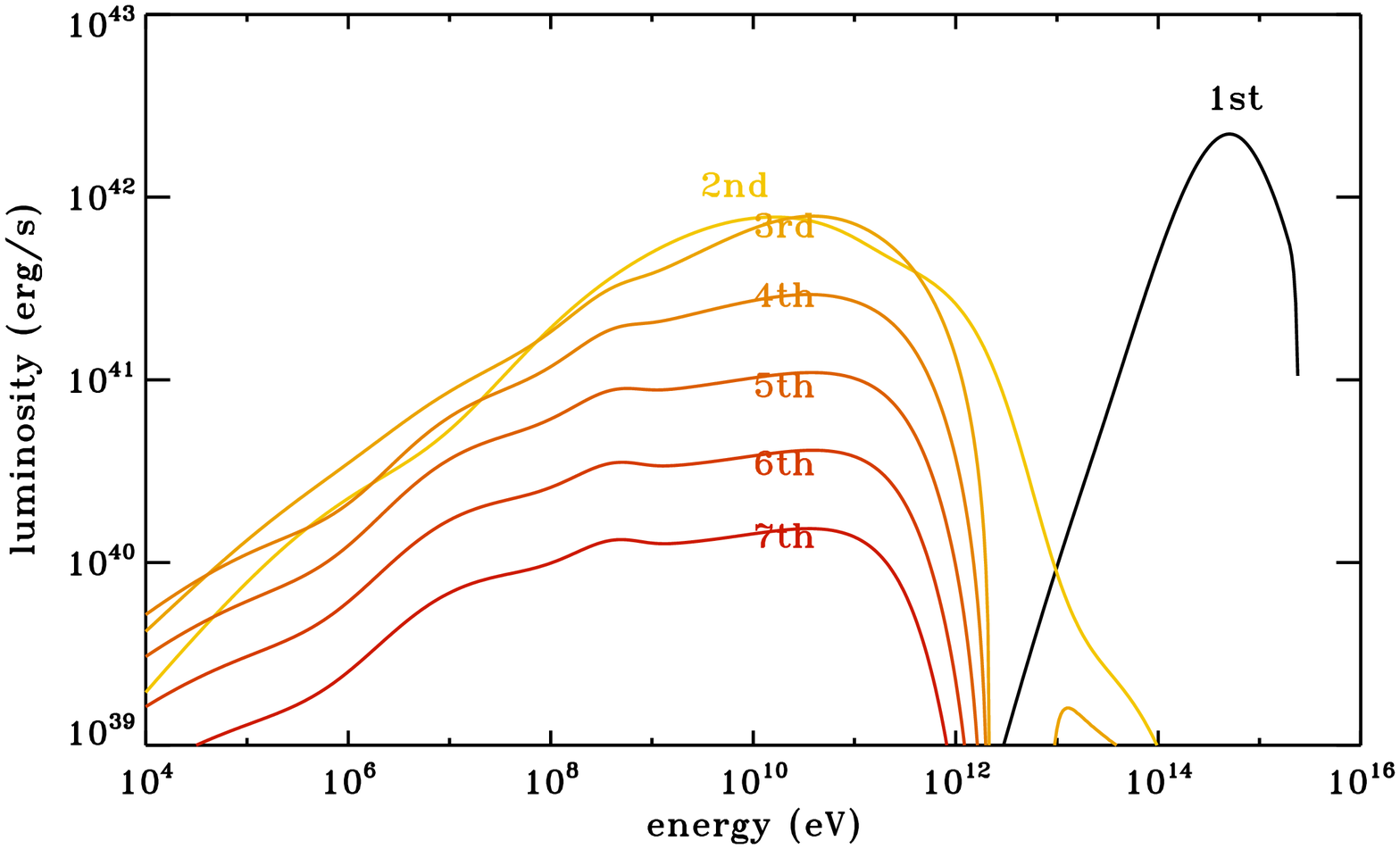}
\caption{The intrinsic EM emissivity decomposed into each generation in the jet's comoving frame. Parameters are the same with those in the upper panel of Fig.~2 in the main text.}
\label{fig:em}
\end{figure}

The generated electrons will produce multiwavelength emission via the synchrotron radiation and the inverse Compton radiation. The total quasi-steady state electron spectrum in the source generated directly by protons and by the first-generation gamma-ray photons are given, respectively, by
\begin{equation}
 \frac{dN_e'}{dE_e'}\Big \vert_p=\left( \frac{d\dot{N}_e'}{dE_e'}\Big \vert_{p\gamma}+\frac{d\dot{N}_e'}{dE_e'}\Big \vert_{\rm BH}\right) \left(t_{\rm jc}^{'-1}+t_{\rm syn}^{'-1} + t_{\rm IC}^{'-1} \right)
 \end{equation}
and
\begin{equation}\label{eq:Nele}
 \frac{dN_e'}{dE_e'}\Big\vert_\gamma= \frac{d\dot{N}_e'}{dE_e'}\Big \vert_{\gamma\gamma} \left(t_{\rm jc}^{'-1}+t_{\rm syn}^{'-1} + t_{\rm IC}^{'-1} \right).
 \end{equation}
These electrons will give rise to the second-generation gamma-ray photons (as well as lower energy emissions) via the synchrotron radiation and IC radiation, i.e.,
\begin{equation}
\frac{d\dot{N}_\gamma}{dE_\gamma}\Big\vert_{\rm 2nd} = \mathcal{F}_{\rm syn}\left\{ \frac{dN_e'}{dE_e'}, B'\right\}+\mathcal{F}_{\rm IC}\left\{ \frac{dN_e'}{dE_e'}, n'(\epsilon')\right\}
\end{equation}
where $\mathcal{F}_{\rm syn}$ and $\mathcal{F}_{\rm IC}$ are operators calculating the synchrotron radiation and the IC radiation which can be found in Ref\cite{Blumenthal70}. A fraction of high-energy gamma-ray photons generated in this step will be absorbed again by the TDE's intense radiation field and produce pairs, as described by Eq.~(\ref{eq:Ngamma}) and Eq.~(\ref{eq:Nele}). The newly generated pairs will give birth to the next-generation gamma rays. Such a cycle will repeat many times (i.e., the so-called electromagnetic cascade) until the energy of the generated photons falls below the threshold for the $\gamma\gamma$ annihilation. We find that the contribution of the 6th or 7th-generation photons are already sufficiently small and hence further cycles can be neglected. The intrinsic EM emissivity in jet's comoving frame from the first seven generations are shown in Fig.~\ref{fig:em}.

Lastly, we sum up the photon flux of each generation, convert to the observer's frame  and take into account the influences of various absorption processes as depicted in the main text.

\section*{\textsl{Fermi}-LAT data analysis}
The \textsl{Fermi} Large Area Telescope (\textsl{Fermi}-LAT) is an imaging, wide field-of-view (FoV) of $\sim 2.4\rm\ sr$, high-energy $\gamma$-ray telescope, covering the energy range from below 20 MeV to more than 300 GeV\cite{Atwood09}. We used Pass 8 SOURCE class events, corresponding to P8R3\_SOURCE\_V2 instrument response functions, and  employed the \textsl{Fermi}-LAT ScienceTools package (fermitools v 1.2.0). We selected a  $17^\circ \times 17^\circ$ Region of Interest (RoI) centered at the AT2019dsg optical position  ($\alpha_{2000}=314.26\rm\ deg,\delta_{2000}=14.20\rm\ deg$), with photon energies from 100 MeV to 800 GeV. We considered the time intervals (230 days) spanning from 2019 April 4 to 2019 November 20, which covers the peak of the optical emission, the UV plateau and the peak of the radio emission. 

\begin{figure}[htbp]
\centering
\includegraphics[width=0.8\columnwidth]{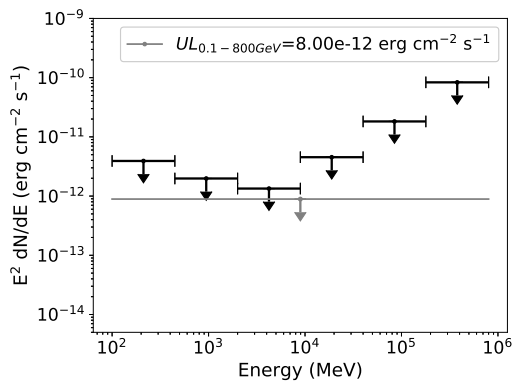}
\caption{$3\sigma$ gamma-ray flux ULs measured by \textsl{Fermi}-LAT from the direction of TDE AT2019dsg over 230 days after the trigger of the TDE (2019 April 4 to 2019 November 20). The gray bar and arrow for total flux UL in 0.1 -- 800\,GeV range, while the black bars and arrows are the differential UL.}
\label{fig:lat_ul}
\end{figure}

We used $gtmketime$ tool to select time interval expressed by (DATA\_QUAL $> 0$) \&\& (LAT\_CONFIG ==1).  To avoid Earth limb 
contamination, we excluded the photons with a zenith angle larger than $90^\circ$.  We binned our data  with a resolution of $0.05^\circ$ per pixel spatially and of 10 logarithmically-spaces bins per energy decade. The background model contains all sources listed in the 4FGL along with the standard diffuse emission background, i.e. the foreground for Galactic diffuse emission ($\rm gll\_iem\_v7.fits$) released and described by the \textsl{Fermi}-LAT collaboration through the \textsl{Fermi} Science Support Center (FSSC)\cite{Acero16} and the background for the spatially isotropic diffuse emission with a spectral shape described by $\rm iso\_P8R3\_SOURCE\_V2\_v01.txt$. The source labelled as $Fermi-$J2113.8+1120 by Ref.\cite{Stein20} is also included in our background model. Since the 4FGL catalog is based on 8\,yr of LAT observations, we found some 4FGL sources with low TS, defined as $\rm{TS}=-2(lnL_0-lnL)$, where $L_0$ is the maximum-likelihood value for null hypothesis and $L$ 
is the maximum-likelihood with the additional source.  We removed the sources with $\rm TS < 6$ (i.e., $<2\sigma$ significance level) from our background model for our short time interval analysis. We can not find any gamma-ray emission  around the position of AT2019dsg, where we test a point-source hypothesis with a power-law spectrum and obtain $\rm TS=0$. We calculate the upper limits (UL) at $3\sigma$ confidence level using the Bayesian methods. The UL for a power-law spectrum with photon power-law index $\Gamma=2$ is $8.0\times10^{-12}\rm\ erg\ cm^{-2}\ s^{-1}$. We also generated the UL within 6 logarithmically space energy bins over 0.1 -- 800\,GeV, as shown in Fig.~\ref{fig:lat_ul} (see also Fig.~2 in the main text).

\bibliography{ms.bib}

\end{document}